\documentclass[traditabstract]{aa}

\usepackage{amsmath}
\usepackage{graphicx}
\usepackage{amssymb}
\usepackage{natbib}
\usepackage{bm}
\usepackage{adjustbox}
\usepackage{txfonts}
\usepackage{xcolor}
\definecolor{yellowgray}{rgb}{0.90, 0.90, 0.2}
\definecolor{bluegray}{rgb}{0.20, 0.60, 0.80}
\definecolor{palered}{rgb}{0.99, 0.40, 0.5}
\definecolor{darkgray}{rgb}{0.35, 0.35, 0.35}
\definecolor{darkgrayb}{rgb}{0.75, 0.75, 0.75}
\definecolor{palegray}{rgb}{0.96, 0.96, 0.96}

\begin{document}

   \title{Solar polarimetry in the K~{\sc i} $D_2$ line: A novel possibility for a stratospheric balloon}

   \author{C. Quintero Noda \inst{1}  \and  G. L. Villanueva\inst{2} \and  Y. Katsukawa\inst{3} \and  S. K. Solanki\inst{4,5} \and  D. Orozco Su\'arez\inst{6} 
   \and  B. Ruiz Cobo\inst{7,8} \and \\  T. Shimizu\inst{1}  \and T. Oba\inst{1,9} \and  M. Kubo\inst{3} \and  T. Anan\inst{10} \and  K. Ichimoto\inst{3,10} \and  Y. Suematsu\inst{3}
          }

   \institute{Institute of Space and Astronautical Science, Japan Aerospace Exploration Agency, Sagamihara, Kanagawa 252-5210, Japan
              \email{carlos@solar.isas.jaxa.jp}     
\and      NASA Goddard Space Flight Center, Planetary Systems Laboratory (Code 693), Greenbelt, MD, USA
\and     National Astronomical Observatory of Japan, 2-21-1 Osawa, Mitaka, Tokyo 181-8588, Japan
\and   Max Planck Institute for Solar System Research, Justus-von-Liebig-Weg 3, D-37077 G\"{o}ttingen, Germany
\and   School of Space Research, Kyung Hee University, Yongin, Gyeonggi, 446-701, Korea
 \and   Instituto de Astrof\'isica de Andaluc\'ia (CSIC), Glorieta de la Astronom\'ia, 18008 Granada, Spain
\and   Instituto de Astrof\'isica de Canarias, E-38200, La Laguna, Tenerife, Spain.
\and   Departamento de Astrof\'isica, Univ. de La Laguna, La Laguna, Tenerife, E-38205, Spain
\and SOKENDAI (The Graduate University for Advanced Studies), Sagamihara, Kanagawa 252–5210, Japan
\and   Kwasan and Hida Observatories, Kyoto University, Kurabashira Kamitakara-cho, Takayama-city, 506-1314 Gifu, Japan 
             }
   \date{Received October, 2017; accepted December, 2017 }

 
  \abstract
{Of the two solar lines,  K~{\sc i} $D_1$ and $D_2$,  almost all attention so far has been devoted to the $D_1$ line, as $D_2$ is severely affected by an O$_2$ atmospheric band. This, however, makes the latter appealing for balloon and space observations from above (most of) the Earth's atmosphere. We estimate the residual effect of the O$_2$ band on the K~{\sc i} $D_2$ line at altitudes typical for stratospheric balloons. Our aim is to study the feasibility of observing the 770~nm window. Specifically, this paper serves as a preparation for the third flight of the Sunrise balloon-borne observatory. The results indicate that the absorption by O$_2$ is still present, albeit much weaker, at the expected balloon altitude. We applied the obtained O$_2$ transmittance to K~{\sc i} $D_2$ synthetic polarimetric spectra and found that in the absence of line-of-sight motions, the residual O$_2$ has a negligible effect on the K~{\sc i} $D_2$ line. On the other hand, for Doppler-shifted K~{\sc i} $D_2$ data, the residual O$_2$ might alter the shape of the Stokes profiles. However, the residual O$_2$ absorption is sufficiently weak at stratospheric levels that it can be divided out if appropriate measurements are made, something that is impossible at ground level. Therefore, for the first time with Sunrise~{\sc iii}, we will be able to perform polarimetric observations of the K~{\sc i} $D_2$ line and, consequently, we will have improved access to the thermodynamics and magnetic properties of the upper photosphere from observations of the K~{\sc i} lines. }


   \maketitle
%

\section{Introduction}

\cite{QuinteroNoda2017b} studied the spectral region at 770~nm containing, among other spectral lines, the K~{\sc i} $D_1$ and $D_2$ lines. The latter is blended with an atmospheric molecular oxygen line belonging to the O$_2$ $A$ band. The authors mentioned that, by observing K~{\sc i}~$D_2$, we can have access to upper photospheric layers (slightly higher than those covered by K~{\sc i} $D_1$), which are located in between the height of formation of traditional photospheric and chromospheric lines such as the infrared Ca~{\sc ii} lines \citep[see Figure~9 of][]{QuinteroNoda2017b}. In this regard, the combination of spectral lines with various heights of formation brings the possibility, for instance, to continuously trace the vertical stratification of the magnetic field or to study the impact of photospheric events on upper atmospheric layers \citep[see, for example, the review of][]{Borrero2015}.   In addition,  the K~{\sc i}~$D_2$ is the most capable line of the doublet for performing quiet Sun polarimetric observations. This is because the K~{\sc i}~$D_2$ line, similar to the Na~{\sc i}~$D_2$ \citep[see, for instance, Figure 1 in][]{Stenflo2000}, produces larger scattering polarization signals, by more than 5 times in the case of the Na~{\sc i} doublet, when observed at various heliocentric angles. Moreover, if we observe the $D_2$ line in combination with the $D_1$ transition we also increase the signal to noise for the polarization signals at the upper photosphere.  This is crucial when  performing inversion
of the Stokes profiles as the amount of information present in an observed
data set is a monotonically increasing function of the number of available spectral lines \citep{AsensioRamos2007}. Finally, we also emphasized in \cite{QuinteroNoda2017b} that the fact that the K~{\sc i} $D_2$ line is completely blocked by the Earth's atmosphere, makes it an appealing candidate for satellite and balloon missions. Therefore, it is useful to estimate the possibilities of observing the potassium lines from a stratospheric balloon, such as the Sunrise solar balloon-borne observatory \citep{Barthol2011,Berkefeld2011,Gandorfer2011}, which has had two successful science flights in 2009 and 2013 \citep{Solanki2010,Solanki2017}. For the above reasons we selected K~{\sc i} $D_1$ and $D_2$ as candidate lines for the next Sunrise flight, i.e. Sunrise {\sc iii}. 

We plan to perform spectropolarimetric observations of those lines, aiming to infer the thermodynamics and properties of the magnetic field in the upper photosphere, around 600~km above the layer where the continuum optical depth is unity at 500~nm. However, this requires that we demonstrate the feasibility of observing the K~{\sc i} $D_2$ line from the stratosphere. Therefore, the main target of the present paper is to assess the influence of the O$_2$ atmospheric absorption on the K~{\sc i} $D_2$ line. 

To this end, we compute the residual effect of the atmospheric O$_2$ in the middle stratosphere, around 35~km, following the characteristics, for example, the launch site, altitude and flight dates, of the two previous flights of the Sunrise mission \citep{Solanki2010,Solanki2017}. We also quantify additional variations of the O$_2$ absorption due to, for instance, changes in the altitude of the balloon or  position of the Sun above the horizon, i.e. different airmass. After computing the O$_2$ transmission profiles for the above scenarios, we apply them to K~{\sc i} $D_2$ synthetic profiles, aiming to measure the effect on the Stokes spectra with special emphasis on the polarimetric signals.  

\section{Method}

\subsection{O$_2$ transmittance}\label{O2trans}

We aim to compute the effect of the oxygen molecular bands on the solar spectra in the 770~nm window. For this purpose, we employed the Planetary Spectrum Generator ({\sc psg}) \citep{Villanueva2017}. This tool\footnote{\url{https://psg.gsfc.nasa.gov}} can be used to generate high-resolution spectra of planetary bodies (e.g. planets, moons, comets, and exoplanets). In addition, the code is able to compute atmospheric transmittances and radiances for various scenarios using the Planetary and Universal Model of Atmospheric Scattering ({\sc pumas}) presented in \cite{Villanueva2015}. This tool performs line-by-line calculations that have been validated and benchmarked with the accurate general line-by-line atmospheric transmittance and radiance model ({\sc genln2}) \citep{Edwards1992genln2}. 

We used {\sc psg} to compute the O$_2$ transmittance spectrum. This spectrum provides information on the amount of solar radiation that is absorbed by the Earth's atmosphere due to the presence of oxygen molecules. In this regard, low transmittance values correspond to large O$_2$ absorption; this absorption is  very prominent at 770~nm, where we can find the O$_2$ $A$ band \citep[e.g.][]{Babcock1948}. The computed wavelength window comprises 20~nm between 755-775~nm with high spectral sampling, similar to that used in \cite{QuinteroNoda2017b}. Additional molecules or aerosols are not considered in this work as O$_2$ is the main contributor to the spectral window of interest. We computed the atmospheric transmittance assuming hydrostatic equilibrium using a vertical profile with 55 layers where the O$_2$ abundance is constant for the range of heights of interest. The O$_2$ molecule considered corresponds to {\sc hitran} \citep{Mcclatchey1973,Rothman2012} number 7 and takes into account all the isotopes of oxygen. We first performed a comparison between the solar atlas of \cite{Delbouille1973} and the solar spectra computed with {\sc psg}. We aim to confirm that the program uses the same wavelength reference that we use later when computing the synthetic polarization signals. The mentioned atlas was observed in the early 70s from the Jungfraujoch International Scientific Station (Switzerland). Therefore, we generated the transmittance profile at 770~nm assuming that we are looking up to the Sun from the Earth \citep[see][for more information]{Villanueva2017}, from the Jungfraujoch observatory located at Switzerland ($7.59^{\circ}$  E, $46.33^{\circ}$  N) at 3500~m above the sea level.  When synthesizing telluric spectra, {\sc psg} accesses the accurate  {\sc nasa-merra2} \citep{Gelaro2017} meteorological database to obtain vertical profiles (from the surface to 70 km) for the observatory site. This provides extremely realistic atmospheric conditions for any site on the planet with a precision of 30 minutes (from 1981 to date) and with a spatial resolution of 1 km (refined employing {\sc usgs-gtopo30} \citep{Gesch1996} topographic information).

\begin{figure}
\begin{center} 
 \includegraphics[trim=+30 0 -10 10,width=8.0cm]{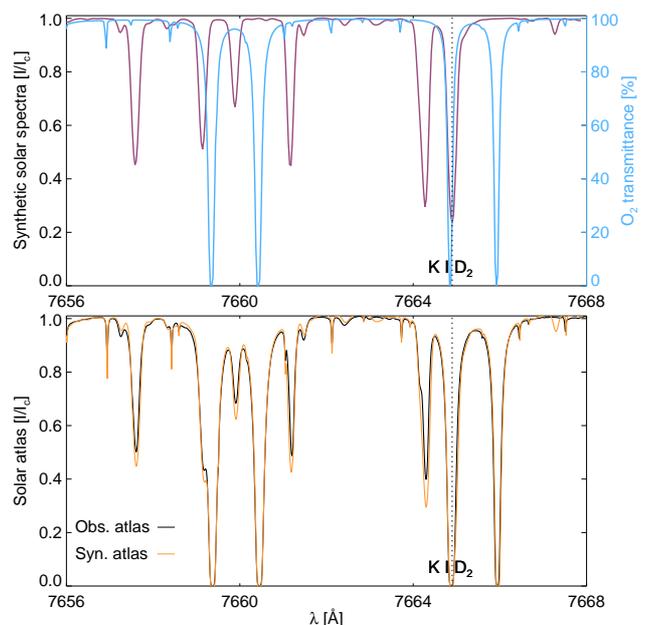}
 \vspace{0.3cm}
 \caption{Top panel: synthetic solar spectra (purple) and telluric O$_2$ transmittance (blue) computed with the {\sc psg} tool. The bottom panel shows  the solar atlas from \citet{Delbouille1973} observed at the Jungfraujoch Scientific Station, Switzerland (black) and the synthetic atlas for the same conditions (orange) generated with the {\sc psg} tool. The vertical dotted line indicates the K~{\sc i} $D_2$ line located at 7664.90~\AA.}
 \label{atlas}
 \end{center}
\end{figure}

We show in the top panel of Figure~\ref{atlas} the solar spectra (without telluric contamination) and the O$_2$ transmittance computed with the {\sc psg} tool. Bottom panel shows a comparison between the observed atlas (black) and the synthesized atlas, i.e. solar spectra and telluric contamination, in orange. Observed and synthetic spectra show very good agreement in amplitude and wavelength, demonstrating the accuracy of the telluric modelling technique. The strong absorption features are dominated by telluric O$_2$ (shown in blue in the top panel), while most of the remaining spectral lines are from solar origin. The O$_2$ telluric absorptions are relatively broad and saturated, and several spectral windows of high transmittance, in which the solar radiation is not blocked by the Earth atmosphere, are present.

\begin{figure}
\begin{center} 
 \includegraphics[trim=-5 0 5 0,width=8.5cm]{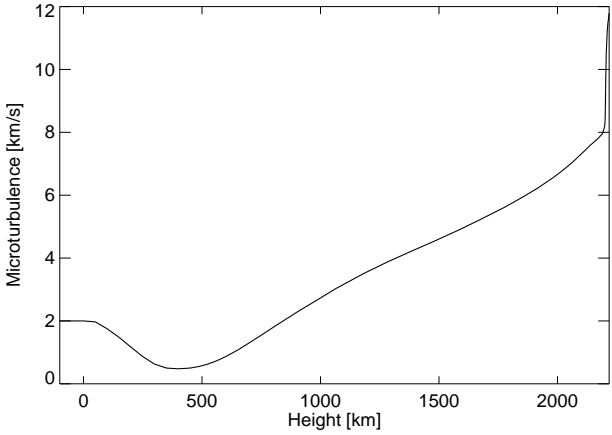}
 \vspace{-0.2cm}
 \caption{Microturbulence stratification used in this work, extracted from \cite{Fontenla1990}.}
 \label{Micro}
 \end{center}
\end{figure}

\begin{figure*}
\begin{center} 
 \includegraphics[trim=-5 0 5 0,width=17.0cm]{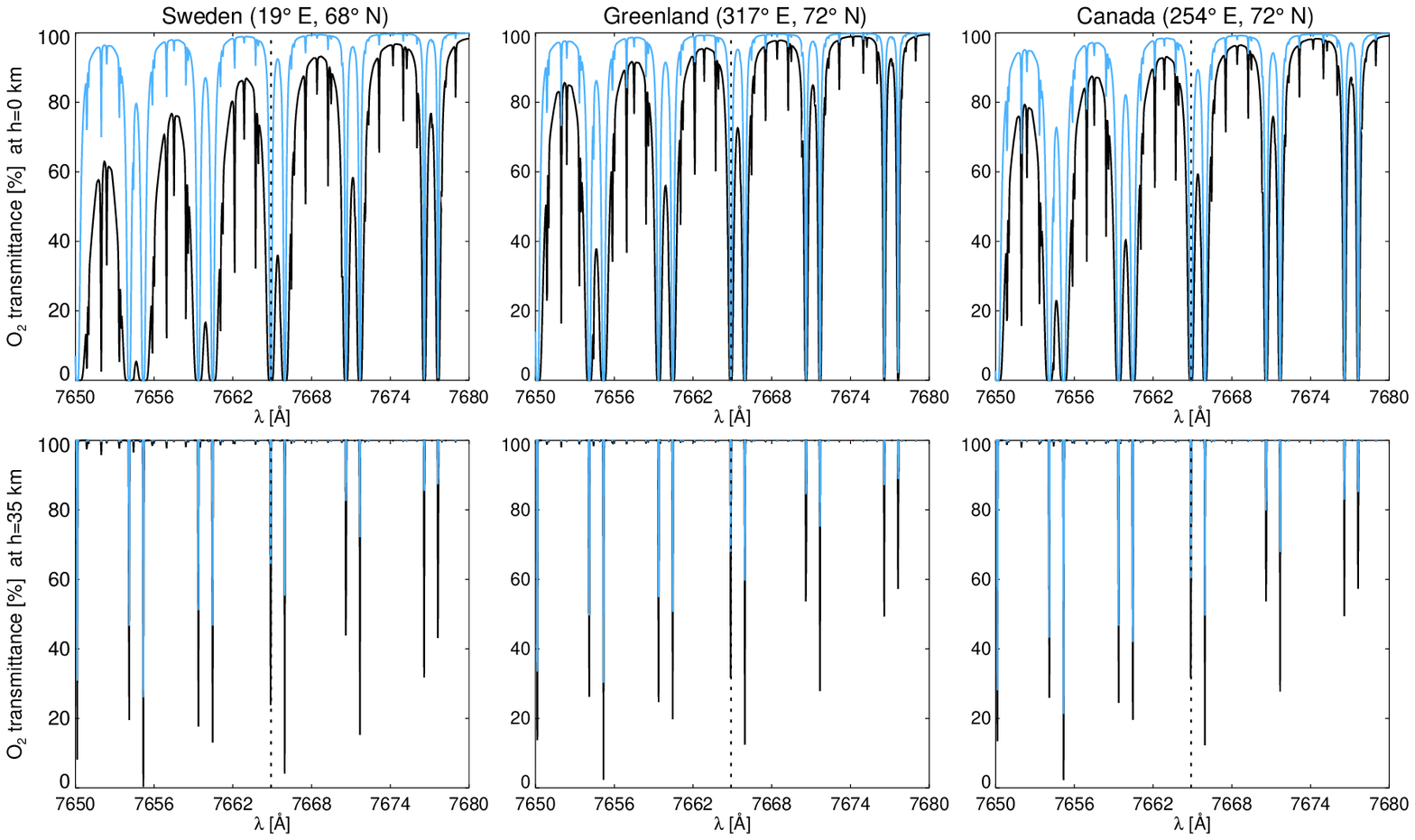}
 \vspace{-0.0cm}
 \caption{Oxygen molecule transmittance in the Earth's northern hemisphere, from left to right, at Sweden, Greenland, and Canada. The top row shows the transmittance at ground level while the bottom row corresponds to the results for an observatory at 35~km above the sea level. Blue designates the transmittance for a reference day time while black corresponds to a reference nocturnal period on 2017 June 15 (see Table~\ref{Coords}).}
 \label{Kiruna}
 \end{center}
\end{figure*}

\subsection{Synthesis of the Stokes parameters}\label{method}

We made use of the {\sc rh} code \citep{Uitenbroek2001,Uitenbroek2003} to synthesize the solar spectra at 770~nm \citep[see Figure 1 in][]{QuinteroNoda2017b}. This region contains the upper photospheric K~{\sc i} $D_2$ line at 7664.90~\AA \ that is heavily blended with an O$_{2}$ line (see the dotted line in Figure~\ref{atlas}). We followed the approach of \cite{QuinteroNoda2017b}, computing the Stokes vector in non-local thermodynamic equilibrium and complete redistribution, as partial redistribution effects are not important for the K~{\sc i} lines \citep{Uitenbroek1992}. We restricted the wavelength coverage to a value of approximately $60$~\AA \ with a spectral sampling of 30~m\AA. The K~{\sc i} atomic model contains 12 levels plus the continuum and was presented in \cite{Bruls1992a} as a simplified version of the comprehensive model. We used the original photoionization and collision values included in the mentioned atomic model and the abundance of K~{\sc i} is taken to be 5.03 \citep[extracted from][]{Asplund2009}.  We employed the semi-empirical FALC model \citep{Fontenla1993}, using the microturbulence presented in Figure~\ref{Micro} and extracted from \cite{Fontenla1990}. We assumed disc centre observations, i.e. $\mu=1$, where $\mu=\cos(\theta)$ and $\theta$ is the heliocentric angle. In addition, we did not include any further spectral degradation and we focussed only on the polarization signals produced by the Zeeman effect; the scattering polarization signals will be studied in the future.

Our objective is to combine the previous synthetic profiles with the transmittance results from {\sc psg}, simulating the path of the light, i.e. generated at the solar atmosphere and perturbed by the Earth's atmosphere. In this regard, we plan to simply multiply the synthetic Stokes parameters by the O$_2$  transmittance profile computed at various observing conditions. We believe this is a correct approach because, given that we always measure a linear combination of Stokes parameters, i.e. $I$+$S$ and $I$-$S$, the effect of the transmission on that combination is the same as when applying the transmission  directly on a given Stokes parameter $S$.

\subsection{Observing conditions} 

We focus on a reference launch site located at the Esrange Space Center, Kiruna, Sweden. This location has been used twice by the Sunrise mission \citep[see][for more details]{Barthol2011,Solanki2017} and implies a flight path that crosses Greenland and Canada. Therefore, we computed the O$_2$ transmittance at these three different locations assuming reference coordinates based on the trajectory followed by the Sunrise balloon in 2009 \citep[see Figure 15 of][]{Barthol2011}. We plan to compare the absorption at ground level and at the stratospheric balloon reference flight height, i.e. around 35~km above sea level. In addition, we studied the transmittance dependence on the day and night cycle and changes in the altitude of the balloon. However, we want to clarify that, at these northern coordinates, the Sun is always visible above the horizon even at night time. Thus, we can observe the Sun at all times during the Sunrise flight. Moreover, as all the computations are performed looking at the Sun, day and night periods simply translate into different solar positions above the horizon.

\section{Results}

\subsection{Launch from Sweden}

We show in Figure~\ref{Kiruna} the O$_2$ transmittance profiles for three reference spatial locations in the northern hemisphere (columns), i.e. Sweden, Greenland, and Canada, computed for day (blue) and night (black) conditions. The latter corresponds to low solar elevation above the horizon but the Sun is still visible. We present the specific details of those locations in Table~\ref{Coords}  where the flight coordinates are estimated from the first Sunrise flight \citep{Barthol2011}. The elevation of the Sun is defined as $\alpha=90^{\circ}$ for the horizon and $\alpha=0^{\circ}$ for the zenith. We distinguish between day and night periods based on the Sun's position above the horizon although the Sun never sets at those latitudes in summer.

The O$_2$ transmittance is always very low at sea level (first row) and is lowest in Sweden. This could be due to the elevation of the Sun above the horizon (see Table~\ref{Coords}) but could be also related to the atmospheric properties on the three geographical locations. In particular, it is probable that the definition of ground level for Greenland corresponds to a higher altitude than that for Sweden or Canada. Still, for all the cases, the O$_2$ band located close to K~{\sc i} $D_2$ (see dots) impedes the detection of the solar line. If we examine a higher altitude, similar to that expected for a stratospheric balloon flight (second row), we find a completely different transmittance spectrum. The first difference is that the transmittance at the K~{\sc i} $D_2$ spectral location is larger than 0, and is up to 60 per cent during the day. But, most importantly, the width of the O$_2$ bands has severely diminished, producing bands of only several m\AA \ wide.  We believe the reasons for this behaviour are the reduction of the atmospheric pressure with height, and consequently the line pressure broadening \citep[for instance,][]{Strong1950}, and that the atmospheric convection at the middle stratosphere is lower than that at sea level. These effects significantly reduce the width of the O$_2$ bands although their absorption is still detectable; the O$_2$ abundance is almost constant up to 150~km.

\begin{table}
\caption{Reference conditions used to simulate a launch from the Esrange Space Center, Kiruna, Sweden (see Figure~\ref{Kiruna}) on 2017 June 15. }\label{Coords} 
\vspace{-0.15cm}
\begin{adjustbox}{width=0.485\textwidth}
  \bgroup
\def\arraystretch{1.25}
\begin{tabular}{lccccccc}
        \hline
Region      &   Coordinates                & Time (UTC) & Period & Sun position  ($\alpha$)    \\
        \hline
Sweden      & (19$^{\circ}$ E, 68$^{\circ}$ N)    & 10:30  & Day   &  44.76$^{\circ}$       \\ 
Sweden      & (19$^{\circ}$ E, 68$^{\circ}$ N)    & 21:30  & Night &  87.36$^{\circ}$       \\ 
Greenland   & (317$^{\circ}$ E, 72$^{\circ}$ N)   & 15:30  & Day   &  48.81$^{\circ}$       \\ 
Greenland   & (317$^{\circ}$ E, 72$^{\circ}$ N)   & 03:30  & Night &  84.50$^{\circ}$       \\ 
Canada      & (254$^{\circ}$ E, 72$^{\circ}$ N)   & 19:00  & Day   &  48.64$^{\circ}$       \\ 
Canada      & (254$^{\circ}$ E, 72$^{\circ}$ N)   & 05:00  & Night &  82.00$^{\circ}$       \\ 
        \hline
  \end{tabular}
  \egroup
\end{adjustbox}    
\end{table}

\begin{figure}
\begin{center} 
 \includegraphics[trim=-7 0 -10 0,width=8.0cm]{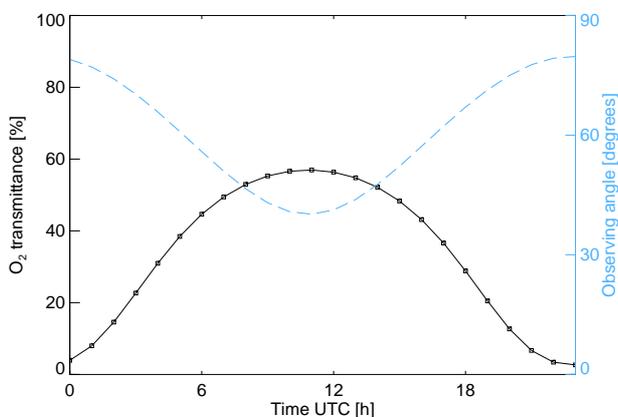}
 \vspace{-0.0cm}
 \caption{Daily variation of the oxygen molecule band transmittance (squares) blended with the K~{\sc i} $D_2$ line at the northern hemisphere (Sweden) for the reference date 2017 June 15. Dashed line indicates the position of the Sun above the horizon, defined as 90$^{\circ}$, while the zenith corresponds to 0$^{\circ}$.}
 \label{Sweden_daynight}
 \end{center}
\end{figure}

Figure~\ref{Kiruna} also reveals the differences between day and night transmittances; the latter has less amplitude. The reason for that is the elevation of the Sun above the horizon. This elevation fluctuates between day and night (see dashed line in Figure~\ref{Sweden_daynight}, we define the zenith as 0$^{\circ}$ and the horizon as 90$^{\circ}$), and consequently, the atmospheric airmass along the  line of sight (LOS) of the observer. In addition, we note that, as the reference launch site in Kiruna is located far north, i.e. 68$^{\circ}$ N, the Sun never reaches a position higher than 50$^{\circ}$ over the horizon even at midday; see the evolution of the observing angle (the zenith corresponds to 0$^{\circ}$) in Fig.~\ref{Sweden_daynight}. Moreover, we also compute the airmass following \cite{Kasten1989}, finding that it closely resembles the pattern displayed by the observing angle (dashed line).

In order to examine the daily variation of the O$_2$ transmittance, we selected a reference spatial location, i.e. Sweden, and we computed the transmittance every hour for a day. We picked the same reference date corresponding to 2017 June 15. Figure~\ref{Sweden_daynight} shows the maximum O$_2$ transmittance (squares) of the band that is blended with the K~{\sc i} $D_2$ solar line for the mentioned period. The transmittance is lower at night (around 10 per cent) while it can reach up to 50$-$60 per cent for a few hours at midday. Importantly, however, it is always above zero. Therefore, at no time is all solar information lost, so that techniques to recover it can be applied. Something that is impossible from the ground, as the transmittance is zero or very close to it (see Figure~\ref{atlas} and \ref{Kiruna}). Still, it seems that the optimum period for observing the K~{\sc i} $D_2$ solar line is around noon rather than at midnight, as the transmittance varies from 60 to 10 per cent depending on the solar position over the horizon.

\begin{figure}
\begin{center} 
 \includegraphics[trim=-13 0 13 15,width=8.0cm]{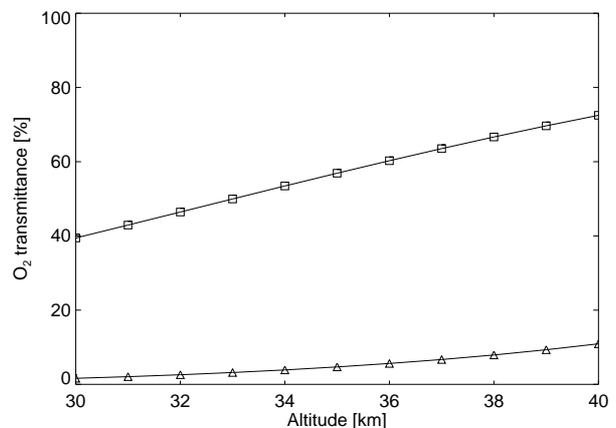}
 \vspace{-0.0cm}
 \caption{Altitude dependence of the oxygen molecule transmittance band blended with the K~{\sc i} $D_2$ line in the northern hemisphere. The geographical location corresponds to Sweden with squares and triangles designating the transmittance at 10:30 and 21:30 UTC, respectively. For more information, see the first two rows in Table~\ref{Coords}.}
 \label{Sweden_height}
 \end{center}
\end{figure}

\begin{figure*}
\begin{center} 
 \includegraphics[trim=5 0 10 0,width=17.0cm]{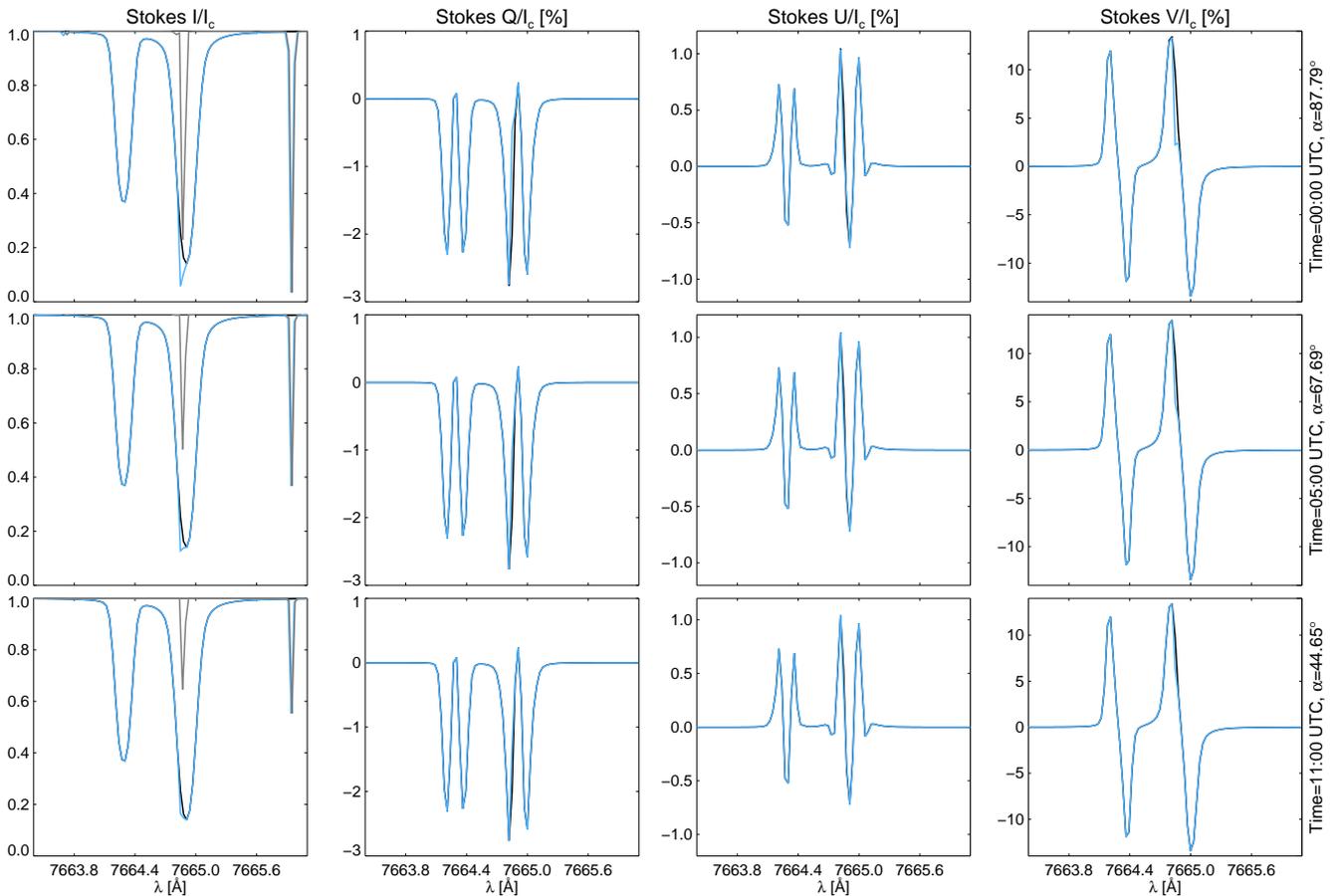}
 \vspace{+0.3cm}
 \caption{Comparison between synthetic profiles without considering the O$_2$ transmittance (black) and after applying it (blue). The grey line indicates the transmittance spectra at 35~km for the reference date 2017 June
15; these spectra are normalized and share the same ordinate axis with the intensity. Each row represents a selected time corresponding to different Sun positions over the horizon; $\alpha=0^{\circ}$ is the zenith. Columns, from left to right, represent the Stokes $I$, $Q$, $U$, $V$ parameters.}
 \label{Kiruna_prof}
 \end{center}
\end{figure*}

Stratospheric balloons change altitude with time over a range of a few kilometres in the course of a day, for example between 34$-$37~km in the case of the Sunrise balloon \citep{Barthol2011}. We represent in Figure~\ref{Sweden_height} the O$_2$ transmittance for the band that is blended with the K~{\sc i} $D_2$ solar line at different altitudes, from 30$-$40~km above sea level, selecting the same date and geographical position used in the bottom panel of the leftmost column of Figure~\ref{Kiruna}. For both, day and night periods, we can see a linear dependence with height; the transmittance values are much larger for the former period, reaching up to 70 per cent if an altitude of 40~km is achieved. However, in spite of this behaviour, we do not detect changes in the width of the O$_2$ band, suggesting that to first order the lines are optically thin at these altitudes, in contrast to when they are observed from the ground. We performed a linear fit of both cases finding that the slope for the day case is roughly 3.3 per cent km$^{-1}$, while, for the night period, it is approximately 0.9 per cent km$^{-1}$. This indicates that we can expect just small variations with time of the O$_2$ transmittance due to altitude changes because the previously recorded altitude fluctuations have an amplitude less than 3~km \citep{Barthol2011}.

\begin{figure*}
\begin{center} 
 \includegraphics[trim=0 0 0 0,width=17.5cm]{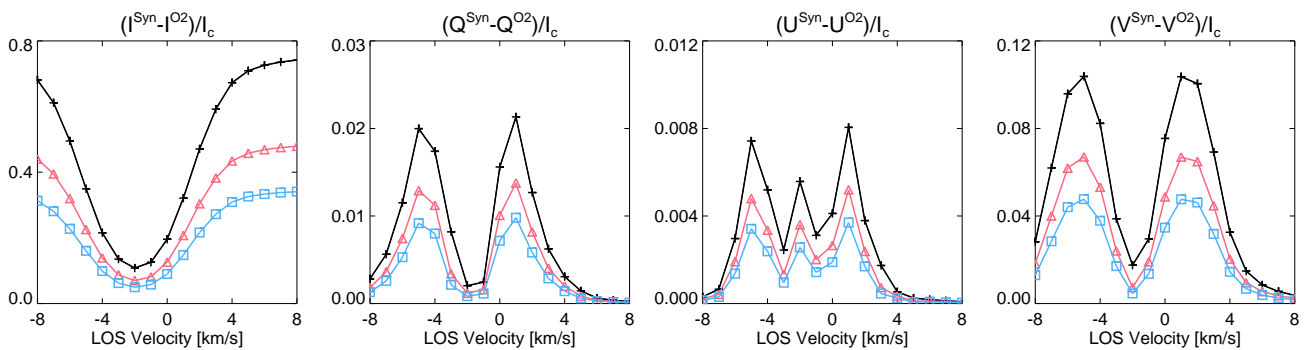}
 \vspace{-0.3cm}
 \caption{Difference between the original synthetic K~{\sc i} $D_2$ Stokes profiles and those affected by the O$_2$ band. We consider different LOS velocities (abscissas) that shift the K~{\sc i} $D_2$ line with respect to the O$_2$ band and we represent the maximum difference for the three selected times (Sun positions) presented in Figure~\ref{Kiruna_prof}, i.e. 00:00 (crosses), 05:00 (triangles), and 11:00 UTC (squares). The velocity sign follows the solar convention, i.e. positive velocities are associated with redshifted profiles and downflows at the solar surface.}
 \label{Doppler_prof}
 \end{center}
\end{figure*}

\section{K~{\sc i} D$_2$ polarimetry at 35 km}

\subsection{Solar lines at rest}

We represent in Figure~\ref{Kiruna_prof} the Stokes profiles for a narrow window centred on K~{\sc i} $D_2$ synthesized following the method explained in Sec.~\ref{method}. The window contains the line of interest and photospheric Fe~{\sc i} line at 7664.30~\AA. We compare the original profiles (black) and the result of multiplying them by the O$_2$ transmittance (blue). We select the same conditions as in previous sections, i.e. Sweden on 2017 June 15, and pick up the transmittance for three selected times (rows) that correspond to different Sun positions (see also Figure~\ref{Sweden_daynight}). In addition, as a reference, we add the O$_2$ transmittance (grey) in the intensity panels (leftmost column). The full width at half maximum of the blended O$_2$ line is less than 70~m\AA \ or, assuming that we scan it with a reference spectral sampling of $\Delta\lambda$=30~m\AA, less than 3 pixels along the spectral direction, and seems to be independent of the Sun elevation. 

This oxygen molecule band reduces the intensity of the K~{\sc i} $D_2$ line by up to 0.1 of the continuum intensity ($I_c$) for the lowest elevation (top row). However, the O$_2$ profile is much narrower than that of the K~{\sc i} $D_2$ line, which means that the general shape of the line is unaltered, mostly the wings. This condition is translated into polarization profiles that barely change, even for low Sun elevations, pointing out that the effect of the O$_2$ band for the presented conditions, i.e. solar lines at rest, is negligible for the polarization profiles. The reason why the effect of the O$_2$ transmittance is negligible is simply because the K~{\sc i} $D_2$ is a strong line and hence strongly saturated.

Finally, we intentionally include the isolated O$_2$ band at the right part of the spectrum to show that we could infer the evolution of the molecular line that is blended with K~{\sc i} $D_2$ if we trace the variations shown by that isolated  O$_2$ band.

\subsection{Wavelength shifted solar lines}

We explained in the previous section that the effect of the oxygen molecule is negligible when the K~{\sc i} $D_2$ line is at rest. However, we want to estimate what happens when we introduce a wavelength shift in the solar spectra with respect to the O$_2$ band. There are various features that can produce this effect, among others, convection in the solar photosphere, solar rotation and Earth's orbital (and partly rotational) motion, or waves in the solar atmosphere.

We show in Figure~\ref{Doppler_prof} the maximum difference between the original synthetic profiles and those affected by the Earth's atmospheric absorption. We select the elevations of the Sun used in the previous section.  On this occasion, the effect of the O$_2$ absorption on the polarization Stokes parameters is not negligible. In the worst scenario, i.e. 00:00 UTC (crosses), we have a maximum difference of 0.02 and 0.01 of $I_c$ in Stokes $Q$ and $U$, respectively, while in Stokes~$V$ we detect larger values up to 0.11 of $I_c$.

\begin{figure}
\begin{center} 
 \includegraphics[trim=0 0 15 15,width=7.5cm]{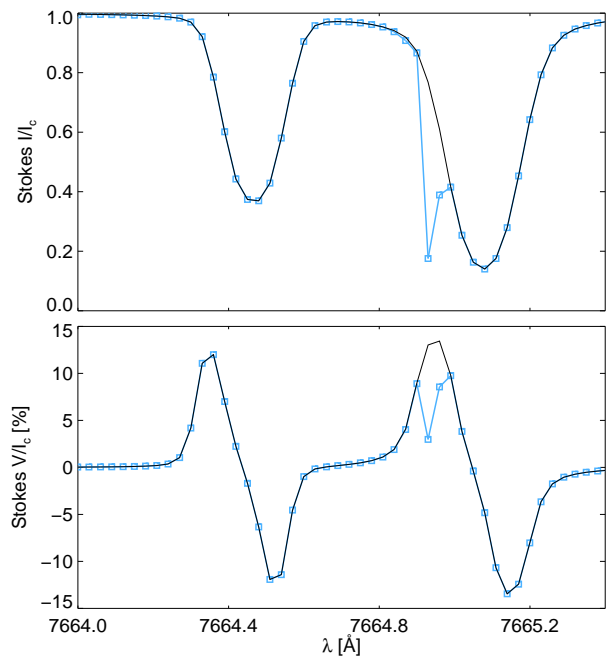}
 \vspace{+0.6cm}
 \caption{Comparison between the original synthetic Stokes $I$ (top) and $V$ (bottom) profiles and those affected by the O$_2$ transmittance. We show the profiles corresponding to the maximum difference (crosses in Figure~\ref{Doppler_prof}) for a given Doppler shift, e.g. 3~km/s.}
 \label{Shift2}
 \end{center}
\end{figure}

 In order to visualize how this affects the Stokes parameters, we plot in Figure~\ref{Shift2} selected profiles that correspond to the Doppler shifts that roughly generate the highest differences, e.g. 3~km/s. Starting with Stokes~$I$ (top), we can see that this corresponds to the case in which the O$_2$ band falls in one of the wings of the K~{\sc i} $D_2$ line, which is compatible with the results presented in Figure~\ref{Doppler_prof}. In Stokes~$V$, we detect the largest deviation when the O$_2$ band modifies one of the lobes. In this case, this is translated into an artificial Stokes~$V$ area and amplitude asymmetry with an amplitude reduction of almost 70 per cent; the original Stokes $V$ lobe diminishes from 0.14 to 0.04 of $I_c$. The same effect occurs for the linear polarization profiles (not presented here) when the oxygen band falls at line core wavelengths as it partially removes the central $\pi$ component. However, we note that those profiles correspond to the worst case scenario, when observing at night with the Sun very close to the horizon (largest airmass). In fact, if we compare the various lines on each panel of Figure~\ref{Doppler_prof}, we detect much lower differences for observations at noon.

\begin{figure*}
\vspace{+0.2cm}
\begin{center} 
 \includegraphics[trim=-5 0 15 15,width=16.0cm]{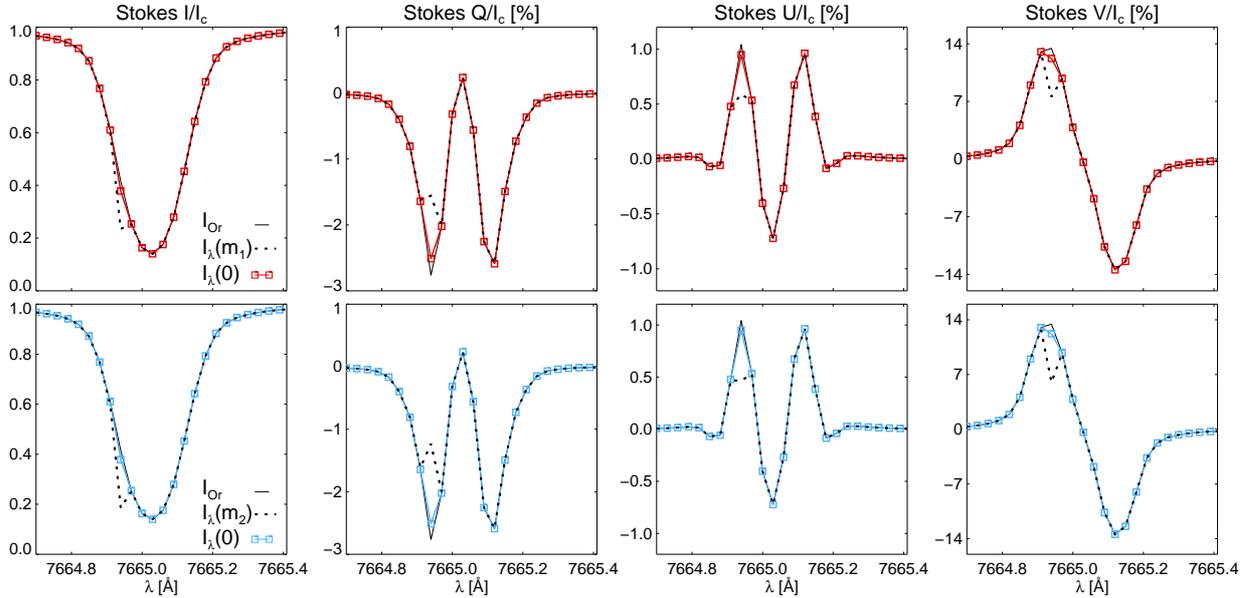}
 \vspace{+0.4cm}
 \caption{Results of applying the telluric correction on the Stokes (from left to right) $I, Q, U, V$ profiles. The top row corresponds to the spectra observed at lower airmass $m_1$, i.e. at 12:00 UTC, while the bottom row designates the spectra at larger airmass $m_2$, i.e. 06:00 UTC.  Solid lines represent the original profile, the dotted lines display the same profile affected by the O$_2$ absorption, and squares show the results obtained from the method explained in Section~\ref{Correc}.}
 \label{Divide}
 \end{center}
\end{figure*}

\section{Telluric correction}\label{Correc}

\cite{Wallace1996} explained that one option for removing the effect of the Earth's telluric absorption consists of observing the spectrum of interest at two different airmasses.  
 Their approach is based on the Beer-Lambert law that relates the attenuation of the light with the properties of the material it is travelling through. In this regard, and focussing on the case of the Earth's atmosphere, we can define the intensity at a given observatory as

\begin{equation}\label{Beer} 
I_{\lambda}(m)=I_{\lambda}(0)\times T_{\lambda}(m),
\end{equation}
where $I_{\lambda}(0)$ is the spectrum produced in the Sun and $T_{\lambda}(m)$ the attenuation induced by the Earth's atmosphere for a given airmass $m$. The attenuation can be described in the present case as
\begin{equation}
T_{\lambda}(m)=e^{-m[\tau_{O_2}(\lambda)]},
\end{equation}
where $\tau_{O_2}(\lambda)$ is the optical depth related to the absorption of the $O_2$ molecular band. As explained in Section~\ref{O2trans}, we assume that the main contribution for the atmospheric attenuation is the oxygen band, although, in general, additional terms should be included for a different spectral range. The airmass $m$, can be defined as $m=\sec(\theta)$, where $\theta$ is the solar zenith angle. 

If we perform observations at two different airmasses, for example observing the Sun at different elevations, we can obtain two spectra $I_{\lambda}(m_1)$ and $I_{\lambda}(m_2)$ whose only difference should be  airmasses $m_1$ and $m_2$. Therefore, we can compute the telluric attenuation by simply dividing the two observations performed at different airmasses as follows:
\begin{equation}\label{Beer3} 
\frac{I_{\lambda}(m_1)}{I_{\lambda}(m_2)}=e^{-\tau_{O_2}(\lambda)(m_1-m_2)},
\end{equation}
as the only unknown of the previous equation is $\tau_{O_2}(\lambda)$. Finally, after determining $\tau_{O_2}(\lambda)$, we can derive $I_{\lambda}(0)$ from Equation~\ref{Beer}. In addition, as $\tau_{O_2}(\lambda)$ does not change strongly on short timescales, in contrast to H$_2$O \citep[see][]{Wallace1996}, we only need to estimate $\tau_{O_2}(\lambda)$ once per day or every two days. We can carry out this estimation while performing, for instance, flat-field calibrations when the Sun elevation is low in the morning and when it is higher around noon; see Figure~\ref{Sweden_daynight}.

In order to reinforce the previous argument, we test the method on synthetic spectropolarimetric spectra. We employ the transmittance presented in Figure~\ref{Sweden_daynight}, where $m_1$ and $m_2$ are the airmasses at 12:00~UTC and 6:00 UTC, respectively. We start from a Stokes vector that is affected by the O$_2$ absorption and also Doppler-shifted with a line-of-sight velocity of 3~km/s. This is because we aim to examine this technique for a similar case as that shown in Figure~\ref{Shift2}. The results of applying the method described before are presented in Figure~\ref{Divide}. The first row shows that we can recover the original Stokes $I$ profile almost perfectly; see the similarities between the squares and the solid line. Moreover, if we replace in Equation~\ref{Beer} $I_{\lambda}$ by the Stokes $Q, U$ and $V$ spectra, we can also recover the polarization signals with high accuracy (see columns in Figure~\ref{Divide}).  This means that we can correct the O$_2$ effect almost perfectly in this ideal case, with unchanging solar spectra and in the absence of noise. The most critical is that the solar spectra remain unchanged between the two measurements. This implies measuring the quiet Sun at the same $\mu$ and same relative velocity (i.e. at the same solar longitude if the relative Sun-Earth velocity is otherwise unchanged), which is approximately the case if we compare observations made at midnight and noon.

\section{Summary and conclusions}

We estimated the O$_2$ transmittance assuming the conditions expected for the Sunrise {\sc iii} balloon-borne mission. In addition, these results are also applicable to any other balloon flight that aims to observe the 770~nm spectral region.

We studied the O$_2$ transmittance dependence on different conditions, such as the altitude and geographical location of the balloon or the Sun elevation. We found that the first condition generates transmittance bands that are very narrow when observing the Sun from the middle stratosphere because of the reduction of atmospheric pressure with height, i.e. the pressure broadening of the spectral lines is much lower \citep[see e.g.][]{Strong1950}, and the lack of convection and associated turbulence at those atmospheric layers. This reduces the effect of the band on solar lines, opening the possibility of observing the solar K~{\sc i} $D_2$ line for the first time. Regarding the geographical location, we did not detect large variations between the three examined locations, although the transmittance is in general slightly lower at Sweden, which could be due to geographical differences but also to the ground level reference altitude (probably higher in Greenland). Concerning the Sun elevation, we found that, although the Sun never sets in summer above the Arctic circle, the transmittance is low during nocturnal periods. This indicates that it is better to observe the K~{\sc i} $D_2$ line close to local noon, when the O$_2$ transmittance is much larger, up to 60 per cent.
 
Later, we studied the effect of the O$_2$ band on synthetic polarimetric spectra, finding that it is significant when we introduce a LOS velocity that shifts the location of the K~{\sc i} $D_2$ line. This is because the residual O$_2$ absorption changes the properties of the Stokes profiles, for example the Stokes $V$ amplitude can be reduced, in the worst case, by up to 70 per cent of its original value. 

We also tested the method presented in \cite{Wallace1996} for removing the telluric contamination from the observed spectra. This technique has been successfully applied on solar, albeit only spectroscopic, observations in the past \citep[e.g.][]{Livingston1991,Wallace1992,Wallace1993,Wallace1996}. The method is based on the Beer-Lambert law and uses two observations taken at different airmasses for recovering the original spectra.  Our results indicate that, for the ideal case studied, we can recover the original solar spectra, including the polarization Stokes profiles, with high accuracy.  Thus, we have found that it is feasible to correct the telluric absorption if we observe (e.g. twice a day) the Sun at different elevations periodically, for instance performing a flat-field calibration, and the airmass of each observation is properly determined. 

We conclude that observing the K~{\sc i} $D_2$ line from a stratospheric balloon is achievable and the spectral properties of the line can be studied. This will allow the scientific exploitation of the data and the discovery of new features in the solar atmosphere, at atmospheric layers that have been scarcely explored with polarimetry.

\section*{Acknowledgements}
We would like to thank the referee for the helpful and constructive comments that helped improve the manuscript. C. Quintero Noda acknowledges the support of the ISAS/JAXA International Top Young Fellowship (ITYF). This work was supported by the funding for the international collaboration mission (SUNRISE-3) of ISAS/JAXA. This project has received funding from the European Research Council (ERC) under the European Union's Horizon 2020 research and innovation programme (grant agreement No. 695075) and has been supported by the BK21 plus programme through the National Research Foundation (NRF) funded by the Ministry of Education of Korea. This work has also been supported by Spanish Ministry of Economy and Competitiveness through the project ESP-2016-77548-C5-1-R.

\bibliographystyle{aa} 
\bibliography{oxygen2.bib} 
\end{document}